\documentclass[fleqn,usenatbib]{mnras}

\usepackage{newtxtext,newtxmath}

\usepackage[T1]{fontenc}
\usepackage{ae,aecompl}

\usepackage{graphicx}	
\usepackage{amsmath}	
\usepackage{amssymb}	

\title[Solar cycle variation in meteor rates]{Solar cycle variation in radar meteor rates}

\author[M. D. Campbell-Brown]{
M. D. Campbell-Brown,$^{1}$\thanks{E-mail: margaret.campbell@uwo.ca}
\\
$^{1}$Department of Physics and Astronomy, University of Western Ontario, London ON N6A 3K7 Canada
}

\date{Accepted 2019 February 21. Received 2019 February 21; in original form 2019 January 10}

\pubyear{2019}

\begin{document}
\label{firstpage}
\pagerange{\pageref{firstpage}--\pageref{lastpage}}
\maketitle

\begin{abstract}

Sixteen years of meteor radar data from the Canadian Meteor Orbit Radar (CMOR) were used to investigate the link between observed meteor rates and both solar and geomagnetic activity. Meteor rates were corrected for transmitter power and receiver noise, and seasonal effects were removed. A strong negative correlation is seen between solar activity, as measured with the 10.7 cm flux, and observed meteor rates. This lends support to the idea that heating in the atmosphere at times of elevated solar activity changes the scale height and therefore the length and maximum brightness of meteors; a larger scale height near solar maximum leads to longer, fainter meteors and therefore lower rates. A weaker negative correlation was observed with geomagnetic activity as measured with the $K$ index; this correlation was still present when solar activity effects were removed. Meteor activity at solar maximum is as much as 30\% lower than at solar minimum, strictly due to observing biases; geomagnetic activity usually affects meteor rates by less than 10 percent. 

\end{abstract}

\begin{keywords}
meteors -- Sun:activity
\end{keywords}



\section{Introduction}

The atmosphere is an excellent detector of meteoroids in the millimeter size range; a single instrument can monitor hundreds of square kilometres of at meteor heights. The effects of the atmosphere cannot therefore be neglected, particularly in the measurement of meteoroid flux. 

Calculating meteoroid fluxes on the Earth involves counting the number of meteors observed over a particular collecting area, and then correcting for observing biases. Meteor patrol radars are particularly suited to calculating fluxes, because they are not limited by weather or daylight and can observe very large collecting areas over long periods of time. The number of echoes detected with the radar depends on the flux of meteoroids, but also on the beam geometry, the transmitting power of the radar, the noise level on the receivers, the efficiency of the detection software and atmospheric effects. 

\subsection{Solar activity effects on meteor rates}
\label{sec:solact}

The effects of the solar cycle on meteor rates were first studied by \citet{bumba1949}, who reviewed visual meteor records from 1844 to 1943, and found that the rates peaked 5 years after maximum solar activity, and were lowest two years after solar minimum. However, it was not possible to rule out changes in cloud cover, for example, and he found similar variations in the number of comets seen and meteorite falls and finds, all of which seem unlikely to be affected by atmospheric variations in the same way as meteors. The diversity of records over the long time interval considered, in addition to the fact that many of the meteor observations were at the times of major showers, also imply that care should be taken with the results.

The next careful analysis of meteor rates and solar activity were done on data from the Omsala radar in Sweden, which ran during the Perseids in 1953 and from 1956 to 1965, and also for a control interval in September in many of those years.  The transmitter power was lowered after 1959, but an experiment was run alternating the power between the two levels every hour in order to compare the rates, and a correction factor was applied to the later data. \citet{lindblad1967} found a dependence on sunspot numbers of the rates of Perseids, with rates at solar minimum a factor of two higher than at solar maximum. The control period, which avoided the dates of major showers, showed a similar trend. The data set included an anomalous increase in the meteor rates in 1963, also observed in Ottawa, Canada \citep{mcintosh1964} and Christchurch, New Zealand \citep{ellyett1964}, which has been attributed to atmospheric heating from a volcanic eruption \citep{kennewell1974}. In his initial work, Lindblad attributed the change in rates to a simple change in atmospheric density; this, however, should lead only to a raising or lowering of ablation heights, with even the height ceiling moving up and down without changing the number of observed meteors. In a later analysis, with data up to 1972,  \citet{lindblad1976} found the same trend, but this time attributed the changing rates to a change in scale height (density gradient); a smaller scale height and steeper density gradient cause meteors to ablate over a shorter distance and reach a higher maximum brightness and therefore electron density. During solar maximum, heating in the atmosphere causes to scale height to increase and meteors of the same mass are fainter. As evidence of this change, Lindblad used simultaneous visual observations and radar observations of Perseids over the observing cycle to determine ablation heights. He found that while the beginning heights were the same over the solar cycle, the ending heights were higher during solar minimum, which implies a smaller scale height. There was a phase offset in the trend, with the lowest meteor rates occurring one to two years before solar maximum and the highest rates 4 to 6 years after solar maximum.

\citet{ellyett1980} pointed out that, in fact, a change in scale height has two effects: one is the change in maximum brightness, and the other the shortening of the trail. They showed that, if meteors are detected randomly along their length, the rate of meteors will depend on the mass distribution index of the population. If there are more bright than faint meteors (a mass distribution index $s$ less than 2), the rate of meteors will fall as the scale height increases, opposite to the effect found by Lindblad. If the mass index is 2, the two effects will cancel each other out. For meteor populations with more faint meteors ($s$>2), the observed rate will decrease with increasing scale height (decreasing density gradient). The temperature of the atmosphere also affects the meteor rate directly through the diffusion rate; \citet{elford1980} looked at the theoretical rate of mass loss as a function of height for an isothermal, winter and summer temperature profile and found strong differences in the trail length and maximum brightness of meteors. He predicted that this would change the rate of meteor echoes by up to 70\%, independent of actual changes of meteor flux on the Earth; more echoes would be seen in the summer months for each hemisphere. He suggested that there may also be a diurnal effect.

Radar data from the Christchurch radar was examined by \citet{prikryl1983}, who looked at the rates of meteors within a few days of ``enhanced" F10.7 flux (which is strongly correlated with sunspot activity). He found a decrease of 10 to 15\% at these times, except during solar minimum, when there was no effect. \citet{ellyett1977} looked at both the Springhill (Ottawa) and Christchurch radars, and found a strong effect in the 1960 - 1961 dataset in Christchurch; the effect was less during the same interval at Springhill, which he attributes to seasonal effects, which enhanced the solar cycle effects in the southern hemisphere and partially cancelled them in the northern hemisphere. There was a stronger negative correlation of sunspot number against meteor rates for the full 1958--1966 Springhill dataset. He looked at the correlation for daytime and nighttime echoes, and found them to be similar. 

Up to the 1980s, it seemed clear that there was a negative correlation between solar activity (measured either with number of sunspots or the flux at 10.7 cm) and meteor rates, possibly with a delay of a few years. However, when a much longer time series from 1958 to 1997 at the Ond\v{r}ejov radar was analysed  \citep{simek1999,pecina1999,simek2002}, a positive correlation was found between meteor rates and sunspot number, with maximum sporadic rates occurring between 1 and 1.5 years after peak solar activity. The echoes analysed were all overdense, with durations greater than 0.4~s, and slightly different phase shifts for echoes lasting longer than 1~s. They attributed previous negative correlations to small numbers (for Bumba's visual study) and anomalously high meteor rates in 1963, near solar minimum, for the Omsala, Springhill and Christchurch radar studies. 

A more recent analysis of visual data by \citet{dubietis2010} considered the rate of sporadic meteors, excluding the antihelion source, in four time intervals through the year free of major showers: January, March, July and September. They required observations to have limiting magnitudes of +5.5 or fainter, and averaged at least 10 observers for each day for the years 1984 to 2006. They found that meteor rates were highest just after maximum solar activity, with a time offset of 1 to 2 years after solar maximum, very similar to the findings of \v{S}imek and Pecina.

The most recent study of solar cycle effects on radar meteor observations did not look at rates, but echo heights; \citet{stober2014} used data from the Canadian Meteor Orbit Radar (CMOR) 38 MHz system taken from 2002 to 2014, and looked at the variation in the peak of the height distribution of observed echoes. In addition to a strong annual variation, they found a change of 4 km between observations taken during solar maximum and solar minimum, consistent with a change in the neutral density of the atmosphere. 

\subsection{Geomagnetic effects}

The effects of geomagnetic activity on meteor rates have also been studied. \citet{lindblad1978} used the Omsala radar dataset and looked particularly at the days around a sector boundary passage by the Earth. He found that, following a sector boundary passage, the global geomagnetic $C_p$ index increased and the meteor rate decreased, both with a delay of 2 or 3 days. The effect seemed to be stronger in data collected at night than data collected during the day. 

The Springhill and Christchurch radar datasets were examined by \citet{prikryl1979,prikryl1983}, this time around the central meridian passage of bright and faint green-corona regions on the Sun. He also found an inverse correlation between the $K_p$ index and the rate of meteors with persistent echoes; in this case, the $K_p$ index reached a minimum and the rate of meteors a maximum a few days after a bright central meridian passage, and the reverse for faint green corona regions. The $K_p$ index is similar to the $C_p$ index, except that $K_p$ varies between 0 and 9; and $C_p$ varies from 0 to 2.5. The effect was still present but less significant when echoes with all durations were considered. 

\subsection{Current work}

It is obvious that, particularly with respect to solar activity, the effects of the atmosphere on meteor rates are not fully understood. This will affect, for example, precise measurements of the flux of meteor showers from year to year. In order to measure the long-term effects of the atmosphere on meteor rates, it is necessary to have equipment which changes as little as possible over at least one solar cycle, and to account for any changes carefully. In the present work, the measured transmitter power and receiver noise of the 38 MHz system of the CMOR radar are used to correct the observed rates, and the data are used to assess the effects of solar and geomagnetic activity on the meteor rates.

\section{Observations}

The Canadian Meteor Orbit Radar (CMOR) consists of three radars operating at 17.45, 29.85 and 38.15 MHz simultaneously. It is located in Tavistock, Ontario, Canada (43.26N, 80.77W), and has been running reliably since 2002. All three frequencies have a single, three-element yagi transmitter and five two-element yagi receivers arranged in an interferometer to locate echoes in the sky; the gain patterns are effectively all-sky. The 29 MHz system has five remote receivers to enable the calculation of trajectories and orbits. This system has been upgraded to run at 15~kW peak power; the other two systems at 17 and 38 MHz transmit approximately 6~kW. The 17~MHz system suffers from noise and terrestrial interference during the day. The 38~MHz system detects the lowest rate of echoes, but it has been running nearly continuously with no changes to its hardware or software. This system is therefore best suited to studying long-term trends. 

The 38~MHz system has an effective limiting magnitude of approximately +8, and overdense echoes are filtered out by the detection software, leaving underdense and transition echoes. It detects an average of 3200 echoes each day, for a total of 16.7 million echoes in the period from January 2002 to October 2018. Of these, about 700,000 are on the echo lines for major meteor showers during their most active times; the others are sporadics or belong to minor showers. 

\begin{figure*}
	\includegraphics[width=8in]{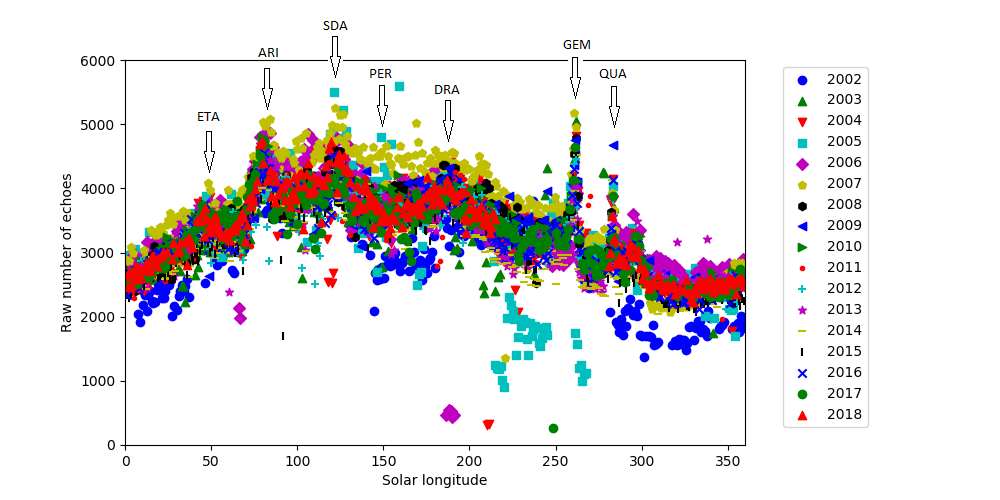}
    \caption{Raw number of echoes on 38 MHz system, 2002 to 2018. Solar longitude is zero on the first day of spring. Major showers are marked. }
    \label{fig:rawnum}
\end{figure*}

The raw number of echoes for each day is shown in figure~\ref{fig:rawnum}. The annual variation is very clear, with a peak in the summer months and the lowest rates in winter. There are also obvious instrumental effects: higher rates for most of 2007, lower rates in parts of 2002, and a time of low rates in 2005, in addition to small outliers in other years.

\subsection{Transmitter Power}

One major influence on the rate of meteor echoes is the power transmitted by the radar. In theory, the radar transmits 6 kW continuously, but small changes in the equipment can cause changes in the power. The power has been measured nearly continuously since 2002; from 2002 to 2011 an in-line power meter which recorded the average power in a text file was used, and since 2009, webcam images of a digital meter were taken at 5 minute intervals. These measurements were processed to find the average transmitted power for each day, excluding times when the power was 0. In some cases, this is because the power was sampled during the cycling of the transmitter, which happens every 30 minutes; in other cases, the transmitter had been turned off for maintenance. 

\begin{figure*}
	\includegraphics[width=7in]{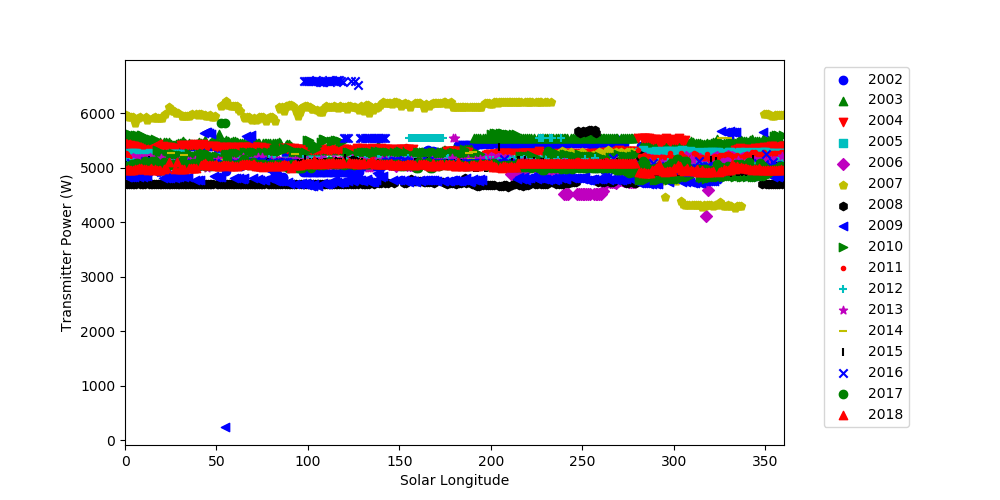}
    \caption{Measured transmitter power for 38 MHz system, 2002 to 2018.}
    \label{fig:38Tx}
\end{figure*}

Figure~\ref{fig:38Tx} shows the transmitter power as a function of time. The power is stable for long stretches, but it does change by about 25\% over the course of sixteen years. In particular, the power was high in 2007, when raw echo rates were also unusually high. 

A correction factor was applied to the meteor rates to remove the effect of changing transmitter power; since the amplitude of an echo is proportional to the square root of the transmitted power, a factor of:

$$C_{Tx} = \sqrt{\frac{P_T}{P_{Tref}}}^{s-1}$$

\noindent was applied to the rates. The reference transmitter power used was 5~kW, but the value chosen is not important, since the goal is to change the rates to a common limiting magnitude. The mass distribution index chosen was $s=2.1$, consistent with the value for sporadics measured for CMOR data in \citet{blaauw2011}.

\subsection{Receiver Noise}

In general, the receiver noise on CMOR is dominated by cosmic sources, such as the galactic centre, the Sun and Cassiopeia A. However, there are instrumental and terrestrial noise sources which occasionally dominate. The noise for each day was calculated using the measured signal-to-noise ratio (SNR) and amplitude of each echo. 

\begin{figure*}
	\includegraphics[width=7in]{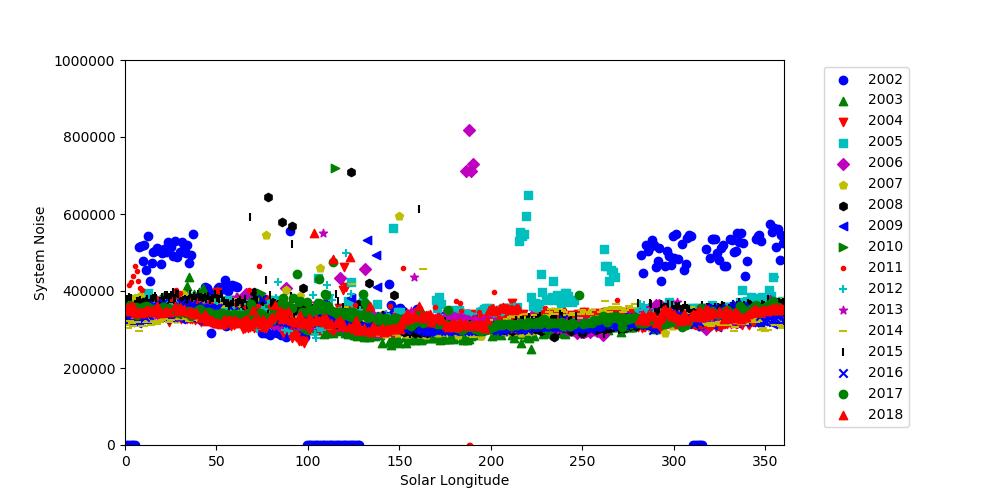}
    \caption{Measured receiver noise in arbitrary linear units for 38 MHz system, 2002 to 2018.}
    \label{fig:38Noise}
\end{figure*}

Figure~\ref{fig:38Noise} shows the 38 MHz receiver noise in linear units. The receiver noise was high in part of 2002 and 2005 (where it accounts for anomalously low observed rates), as well as a few days in other years. The rates were corrected in a similar way to the transmitter power corrections, with:

$$C_{N} = \sqrt{\frac{P_{Nref}}{P_N}}^{s-1}.$$

\noindent The reference noise power was 330000, a typical value; again, the value used does not affect the relative rates. 

\subsection{Signal-to-Noise Ratio}

\begin{figure*}
	\includegraphics[width=7in]{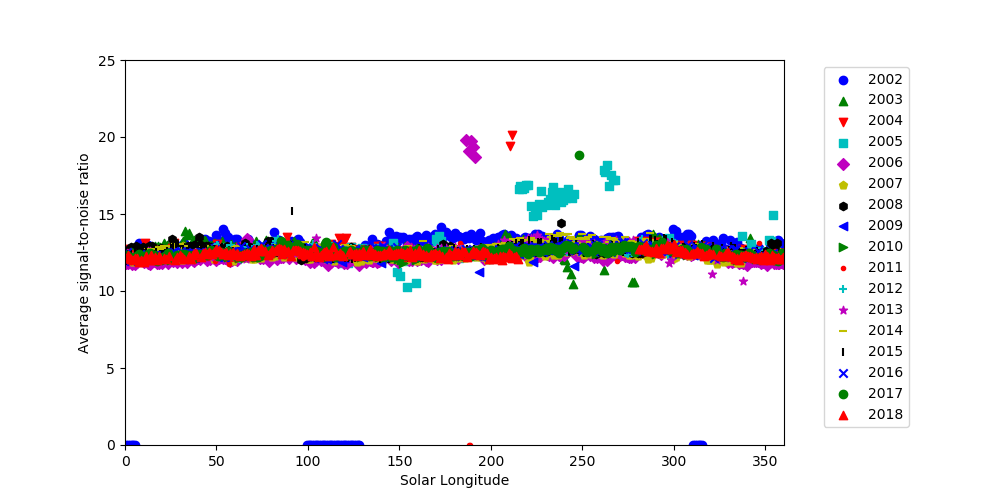}
    \caption{Average signal-to-noise ratio for 38 MHz system, in dB, 2002 to 2018.}
    \label{fig:38SNR}
\end{figure*}

The higher noise in 2005 accounted for some of the drop in rates, but even with the correction for noise level, some of the rates are still anomalously low. The average signal-to-noise ratio (in dB) for echoes was calculated for each day; the results are shown in figure~\ref{fig:38SNR}. 

The typical value is approximately 12.5, but some days have considerably higher averages; this was traced to bad antenna phases, which meant that only very strong echoes had reliable interferometry. No attempt was made to correct the rates for this effect: a limit of 14 was chosen, and all days with an average SNR greater than this were dropped from the analysis. About 50 days were affected. Note that since the average SNR was not used to correct the rates, it does not matter whether linear or logarithmic units are used. 

\section{Results}

\begin{figure*}
	\includegraphics[width=7in]{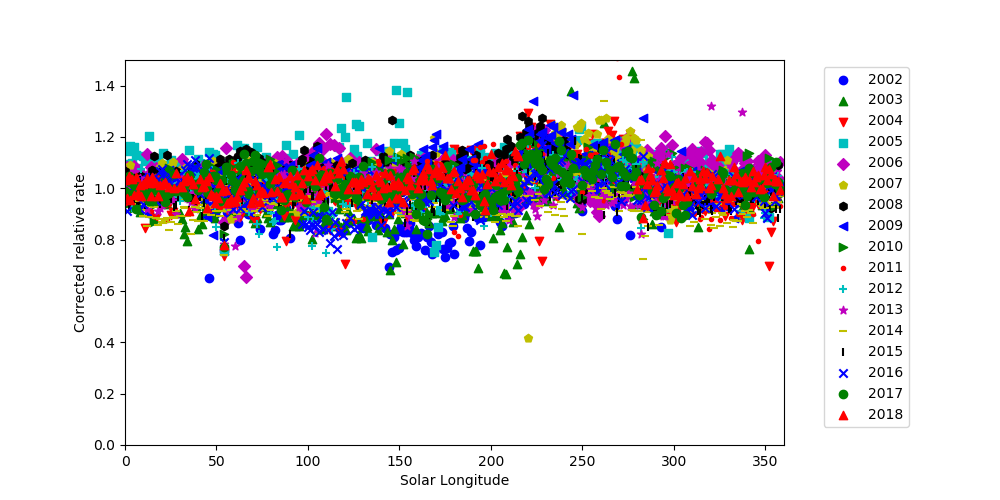}
    \caption{Echo rate relative to the average for each degree solar longitude, corrected for transmitter power and receiver noise, for 38 MHz system, 2002 to 2018.}
    \label{fig:38CorRates}
\end{figure*}

The raw rates for each day were corrected for transmitter power and receiver noise. There remains the seasonal variation of the meteoroid environment, which must be removed before a correlation can be sought between the rates and solar and geomagnetic activity. The average corrected rate was found for each degree solar longitude, and the rate for each day was then divided by this number. This removes any seasonally repeating changes in the meteor rate, including changes in the sporadic sources and in showers which repeat every year. The results, scaled to unity, are shown in figure~\ref{fig:38CorRates}. 

Some deviations from the average are still visible, but most of the obvious instrumental effects have been removed. No showers remain visible in the corrected data.

\subsection{Solar effects}

\begin{figure*}
	\includegraphics[width=8in]{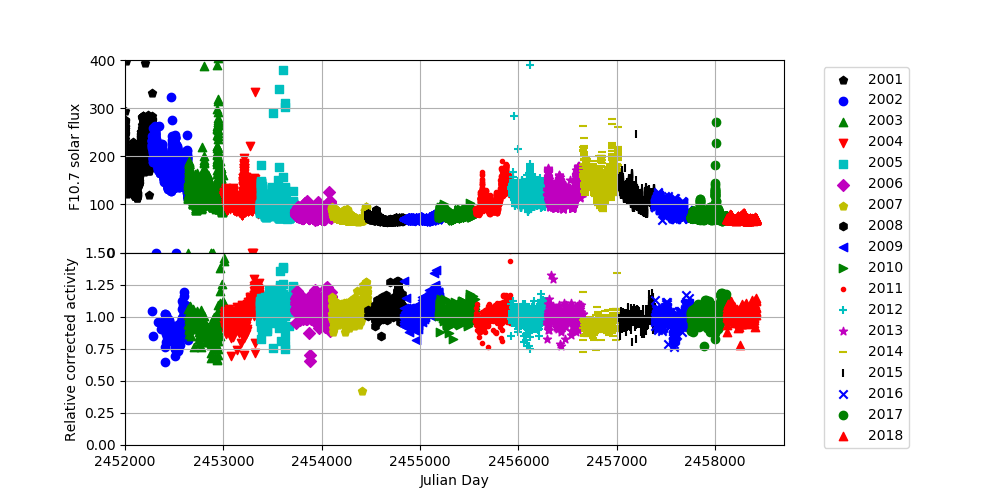}
    \caption{Solar 10.7~cm flux (F10.7 index, in units of $10^{-22}$~Wm$^{-2}$Hz$^{-1}$) and relative meteor echo rate (corrected for transmitter power and receiver noise) on 38 MHz system as a function of julian day.}
    \label{fig:f107andAct}
\end{figure*}

The data can now be examined for atmospheric effects. The flux of 10.7~cm radiation from the Sun is measured three times a day\footnote{http://www.spaceweather.gc.ca/solarflux/sx-en.php}; these three values were averaged to get daily numbers. The radar data covers an interval from just after solar maximum in 2002, through a solar minimum in 2008 and 2009, to a second, lower maximum in 2014. Figure~\ref{fig:f107andAct} shows the solar activity and relative corrected meteor rates plotted against julian day. There is a suggestion that rates are lowest at times of highest solar activity. 

\begin{figure*}
	\includegraphics[width=8in]{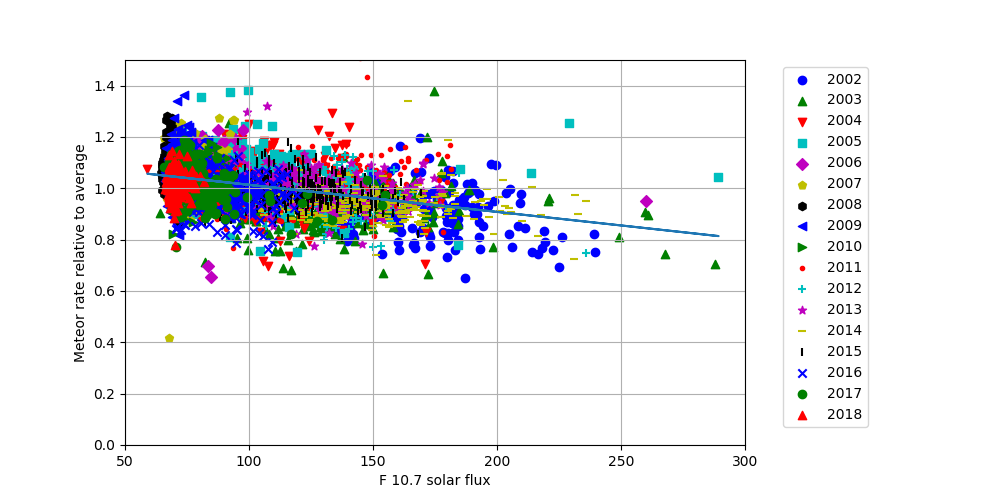}
    \caption{Relative echo rate as a function of solar 10.7~cm flux (F10.7 index, in units of $10^{-22}$~Wm$^{-2}$Hz$^{-1}$) on 38 MHz system.}
    \label{fig:107cor}
\end{figure*}

To quantify the effect of the solar cycle on meteor rates, we can look at the correlation. Figure~\ref{fig:107cor} shows the trend between relative rate and solar activity; even with significant scatter, there is a clear inverse dependence of rate on solar activity. The R-squared of a linear fit is only 0.17. We tried binning the data in 10-day intervals, to see if that would reduce the noise and improve the fit, but the R-squared was smaller for the binned data, 0.12. Considering earlier work, we also introduced a phase offset for the solar activity data, ranging up to 2 years: the results are shown in table~\ref{table:phases}. The fit is slightly worse for offsets of up to one year, and noticeably worse at 2 years. 

With zero phase, the activity varies with solar activity (in units of $10^{-22}$~Wm$^{-2}$Hz$^{-1}$) as:

\begin{equation}
    \label{eq:SolarCor}
    N_{cor} = \frac{N}{-0.001205 * F_{10.7} + 1.0759}.
\end{equation}

\noindent Fluxes at solar maximum to are about 30\% lower than at solar minimum. Here, the correction is set so that at a F10.7 flux of 63$\times10^{-22}$~Wm$^{-2}$ Hz$^{-1}$, the smallest flux recorded in our interval, has a correction of 1, and higher meteor rates will be produced at higher solar fluxes. 

\begin{table}
\caption{ Goodness of fit of solar activity (F10.7 flux) and relative meteor rate for phase offset in years} 

\label{table:phases}
\begin{tabular}{lc}
\hline
Phase (years) & R-squared\\
\hline
0.00 & 0.17 \\
0.25 & 0.13 \\
0.50 & 0.15 \\
0.75 & 0.15 \\
1.00 & 0.11 \\
1.25 & 0.10 \\
1.50 & 0.12 \\
1.75 & 0.10 \\
2.00 & 0.06 \\
\hline
\end{tabular}
\end{table}

To look at short-term effects on the atmosphere, the activity of daytime and nighttime meteors were considered separately. Meteors between 03 and 09 UT (10 pm to 4 am local time) were used for the nighttime analysis, and from 15 to 21 UT (10 am to 4 pm local) were included in the daytime analysis. All meteors had the same correction for transmitter power, but the noise (which varies diurnally and is 40\% higher during the day) was separately calculated for each subset of meteors. 

The same analysis procedures were followed as in the case of the full day dataset. The daytime echoes had a very slightly higher correlation and steeper slope (slope = -0.001586; intercept = 1.1609; r-squared = 0.18), while the nighttime echoes had a lower correlation and shallower slope (slope = -0.001111; intercept = 1.1117; r-squared = 0.1096). 

\subsection{Geomagnetic effects}

The same relative meteor rates, corrected for instrumental effects and relative to the average for a particular solar longitude, were used to look for variations due to geomagnetic effects. To maintain consistency with previous work, we looked at the dependence of rates on the global geomagnetic index $K_p$ (ftp.gfz-potsdam.de/pub/home/obs/kp-ap/), but we also used the local K index from Ottawa, the closest geomagnetic observatory (ftp.geolab.nrcan.gc.ca/pub/forecast/k\_indices/). The unitless K index varies between 0 and 8, with two steps between each, so the series [0o, 1+, 1-, 1o,...] was converted to [0, 0.3, 0.7, 1.0,...]. 

There was a weak inverse correlation of meteor rates and Local geomagnetic index. Phase offsets of 0 to 5 days were tested, and the strongest correlation was found with zero phase offset. Both the maximum effect and the correlation were lower than for the solar activity variation case. 

\begin{figure*}
	\includegraphics[width=7in]{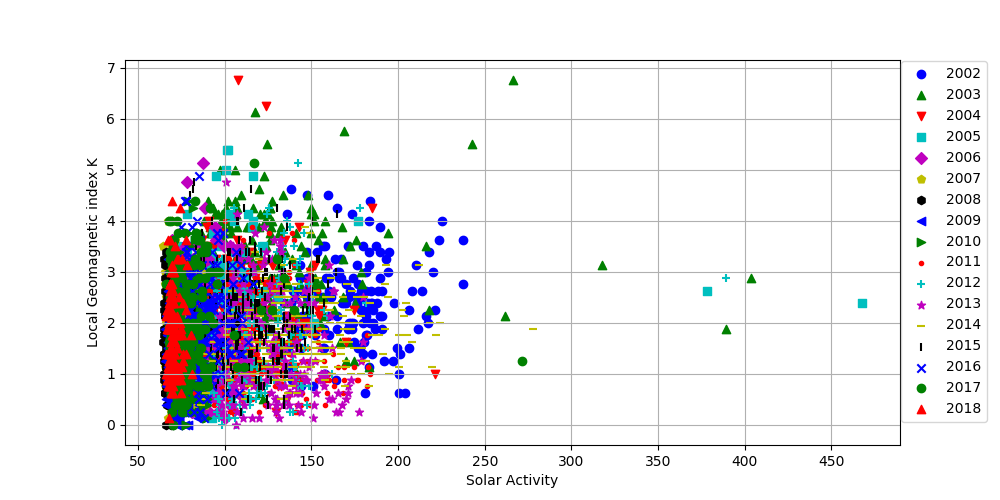}
    \caption{Local (OTT) geomagnetic index and solar activity from the 10.7~cm flux, in units of $10^{-22}$~Wm$^{-2}$Hz$^{-1}$.}
    \label{fig:K107}
\end{figure*}

Solar activity and the geomagnetic index are not strongly correlated, but the K index tends to be higher at times of elevated solar activity (Fig.~\ref{fig:K107}). To rule out solar activity being responsible for the observed variation with geomagnetic index, all the meteor rates were corrected using the solar correction derived in the previous section (Eq.~\ref{eq:SolarCor}). 

\begin{figure*}
	\includegraphics[width=8in]{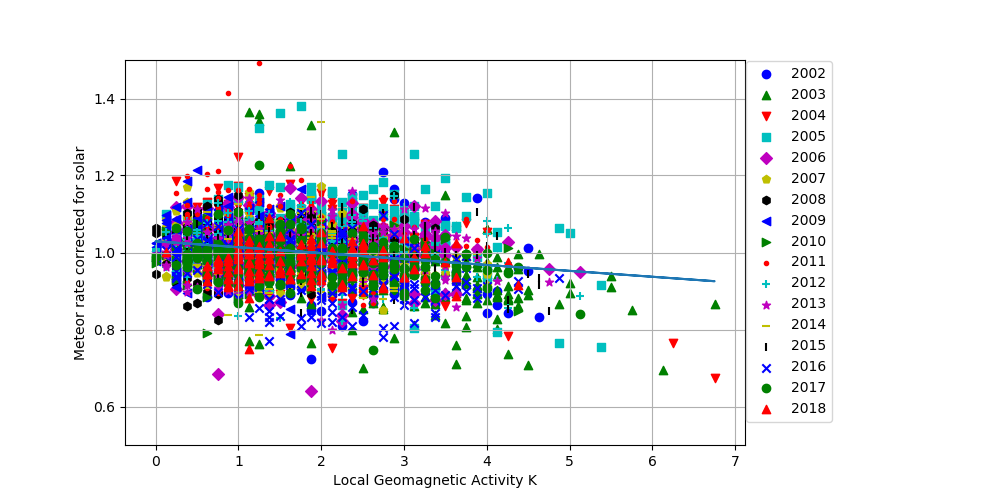}
    \caption{Relative meteor activity, corrected for solar variation as well as instrumental and seasonal effects, as a function of Local (OTT) geomagnetic index}
    \label{fig:KvsSolarCor}
\end{figure*}

The results are shown in Fig.~\ref{fig:KvsSolarCor}, with the line of best fit. The r-squared value is only 0.028, indicating a very weak correlation, but meteor rates are slightly lower when the geomagnetic index is high. The slope is weaker than before the data were corrected for solar activity (-0.015 compared to -0.022), but a slight trend is still present. Even at high K values, however, the straight line fit gives a correction of only a few percent; the correction factor is:

\begin{equation}
    \label{eq:GeomagCor}
    N_{cor} = \frac{N}{-0.01535 * K + 1.000}.    
\end{equation}

\noindent Here the lowest K index is 0, so the correction is 1 at $K$=0 and increases with increasing index, to a maximum of 16\% at the highest $K$ of 9. Very few days of data are affected with the highest K index, and most will have corrections less than 10\%.

\section{Discussion}

There is a clear anti-correlation between solar activity and meteor rates on the 38 Mhz radar. The fact that there is no delay between increasing solar activity and decreasing meteor rates implies that the atmosphere is reacting relatively quickly to solar heating. The fact that the decrease in rates is more pronounced for meteors recorded during the day could imply that density gradient changes occur over short time intervals, but other factors might be responsible. For example, if the ionosphere density and vertical extent are larger at times of elevated solar activity, Faraday rotation may be increased and meteor rates reduced during the day because of the change in the polarization of the echoes. This temporal effect should be investigated in any shower analysis which looks at the variation of rates over periods of less than a day. 

The same analysis was performed on rates from the 29 MHz CMOR radar, using the measured power and noise for that system. There was no trend of the rates on solar activity, and visual inspection of the rates showed systematic increases and decreases in the rates which must be the result of uncorrected changes in the radar sensitivity. The 29 MHz system has undergone many upgrades during its lifetime, and the parameters used to correct the data here are inadequate to describe the sensitivity changes. Assuming the change in meteor rates with solar activity are caused by a change in the scale height of the atmosphere, the rates should change in the same way at any frequency. If some other effect is involved, such as Faraday rotation, there may be a wavelength dependence. Similarly, meteor brightness should be affected in the same way as the maximum ionization, and therefore optical rates should investigated for solar effects. It may be more difficult, however, to correct for changing observing conditions.

Meteor rates should certainly be corrected for solar activity; the effects of geomagnetic activity are less pronounced, but should be kept in mind. It is unclear how the atmosphere at meteor heights reacts to different levels of geomagnetic activity, and so whether the results here are generally applicable. 

\section{Conclusions}

Solar and geomagnetic activity affect the atmosphere, and therefore the observed rate of meteors. Solar activity can depress the observed meteor flux by 30\% at the highest solar flux, and should be taken into account when looking at shower activity in different years. Geomagnetic activity has a smaller effect, but should be considered in comparing rates when the uncertainties are low. 

The fact that the dependence of rates on solar activity is slightly higher during the day implies that either the atmosphere responds very rapidly to changes in solar flux, or that the dependence of meteor rates on solar activity is due to some combination of atmospheric density changes and short-lived effects, such as changes in the ionosphere. Further observations at different frequencies could help to shed light on this, and determine whether a diurnal correction is dependent on frequency. The effects of Faraday rotation due to changes in the ionosphere on meteor rates should be further investigated by explicitly looking at the electron density in the D-region; the role of sporadic E layer interference, which is negligible at 38 MHz but very important at lower frequencies, should also be examined.

When looking at long-term atmospheric effects, it is very important that instrumental effects be carefully tracked and corrected. This study emphasizes the need for long-term monitoring with unchanging equipment, since corrections for significant changes cannot always be successfully made. 

\section*{Acknowledgements}

Funding  for  this  work  was  provided  through  NASA  cooperative agreement 80NSSSC18M0046 and the Natural Sciences and Engineering Research Council of Canada (Grant no. RGPIN-2018-05474).




\bibliographystyle{mnras}
\input{solarvar.bbl}
\bibliography{solarvar} 

\begin{thebibliography}{}
\makeatletter
\relax
\def\mn@urlcharsother{\let\do\@makeother \do\$\do\&\do\#\do\^\do\_\do\%\do\~}
\def\mn@doi{\begingroup\mn@urlcharsother \@ifnextchar [ {\mn@doi@}
  {\mn@doi@[]}}
\def\mn@doi@[#1]#2{\def\@tempa{#1}\ifx\@tempa\@empty \href
  {http://dx.doi.org/#2} {doi:#2}\else \href {http://dx.doi.org/#2} {#1}\fi
  \endgroup}
\def\mn@eprint#1#2{\mn@eprint@#1:#2::\@nil}
\def\mn@eprint@arXiv#1{\href {http://arxiv.org/abs/#1} {{\tt arXiv:#1}}}
\def\mn@eprint@dblp#1{\href {http://dblp.uni-trier.de/rec/bibtex/#1.xml}
  {dblp:#1}}
\def\mn@eprint@#1:#2:#3:#4\@nil{\def\@tempa {#1}\def\@tempb {#2}\def\@tempc
  {#3}\ifx \@tempc \@empty \let \@tempc \@tempb \let \@tempb \@tempa \fi \ifx
  \@tempb \@empty \def\@tempb {arXiv}\fi \@ifundefined
  {mn@eprint@\@tempb}{\@tempb:\@tempc}{\expandafter \expandafter \csname
  mn@eprint@\@tempb\endcsname \expandafter{\@tempc}}}

\bibitem[\protect\citeauthoryear{{Blaauw}, {Campbell-Brown}  \&
  {Weryk}}{{Blaauw} et~al.}{2011}]{blaauw2011}
{Blaauw} R.~C.,  {Campbell-Brown} M.~D.,   {Weryk} R.~J.,  2011, \mn@doi
  [\mnras] {10.1111/j.1365-2966.2010.18038.x}, \href
  {http://adsabs.harvard.edu/abs/2011MNRAS.412.2033B} {412, 2033}

\bibitem[\protect\citeauthoryear{{Bumba}}{{Bumba}}{1949}]{bumba1949}
{Bumba} V.,  1949, Bulletin of the Astronomical Institutes of Czechoslovakia,
  \href {http://adsabs.harvard.edu/abs/1949BAICz...1...93B} {1, 93}

\bibitem[\protect\citeauthoryear{{Dubietis} \& {Arlt}}{{Dubietis} \&
  {Arlt}}{2010}]{dubietis2010}
{Dubietis} A.,  {Arlt} R.,  2010, \mn@doi [Earth Moon and Planets]
  {10.1007/s11038-010-9351-6}, \href
  {http://adsabs.harvard.edu/abs/2010EM%26P..106..105D} {106, 105}

\bibitem[\protect\citeauthoryear{{Elford}}{{Elford}}{1980}]{elford1980}
{Elford} W.~G.,  1980, in {Halliday} I.,  {McIntosh} B.~A.,  eds,  IAU
  Symposium Vol. 90, Solid Particles in the Solar System. pp 101--104

\bibitem[\protect\citeauthoryear{{Ellyett}}{{Ellyett}}{1977}]{ellyett1977}
{Ellyett} C.,  1977, \mn@doi [\jgr] {10.1029/JA082i010p01455}, \href
  {http://adsabs.harvard.edu/abs/1977JGR....82.1455E} {82, 1455}

\bibitem[\protect\citeauthoryear{{Ellyett} \& {Keay}}{{Ellyett} \&
  {Keay}}{1964}]{ellyett1964}
{Ellyett} C.~D.,  {Keay} C.~S.~L.,  1964, \mn@doi [Science]
  {10.1126/science.146.3650.1458}, \href
  {http://adsabs.harvard.edu/abs/1964Sci...146.1458E} {146, 1458}

\bibitem[\protect\citeauthoryear{{Ellyett} \& {Kennewell}}{{Ellyett} \&
  {Kennewell}}{1980}]{ellyett1980}
{Ellyett} C.~D.,  {Kennewell} J.~A.,  1980, \mn@doi [\nat] {10.1038/287521a0},
  \href {http://adsabs.harvard.edu/abs/1980Natur.287..521E} {287, 521}

\bibitem[\protect\citeauthoryear{{Kennewell} \& {Ellyett}}{{Kennewell} \&
  {Ellyett}}{1974}]{kennewell1974}
{Kennewell} J.~A.,  {Ellyett} C.~D.,  1974, \mn@doi [Science]
  {10.1126/science.186.4161.355}, \href
  {http://adsabs.harvard.edu/abs/1974Sci...186..355K} {186, 355}

\bibitem[\protect\citeauthoryear{{Lindblad}}{{Lindblad}}{1967}]{lindblad1967}
{Lindblad} B.~A.,  1967, Meddelanden fran Lunds Astronomiska Observatorium
  Serie I, \href {http://adsabs.harvard.edu/abs/1967MeLuF.226.1029L} {226,
  1029}

\bibitem[\protect\citeauthoryear{{Lindblad}}{{Lindblad}}{1976}]{lindblad1976}
{Lindblad} B.~A.,  1976, \mn@doi [\nat] {10.1038/259099b0}, \href
  {http://adsabs.harvard.edu/abs/1976Natur.259...99L} {259, 99}

\bibitem[\protect\citeauthoryear{{Lindblad}}{{Lindblad}}{1978}]{lindblad1978}
{Lindblad} B.~A.,  1978, \mn@doi [\nat] {10.1038/273732a0}, \href
  {http://adsabs.harvard.edu/abs/1978Natur.273..732L} {273, 732}

\bibitem[\protect\citeauthoryear{{McIntosh} \& {Millman}}{{McIntosh} \&
  {Millman}}{1964}]{mcintosh1964}
{McIntosh} B.~A.,  {Millman} P.~M.,  1964, \mn@doi [Science]
  {10.1126/science.146.3650.1457}, \href
  {http://adsabs.harvard.edu/abs/1964Sci...146.1457M} {146, 1457}

\bibitem[\protect\citeauthoryear{{Pecina} \& {{\v S}imek}}{{Pecina} \& {{\v
  S}imek}}{1999}]{pecina1999}
{Pecina} P.,  {{\v S}imek} M.,  1999, \aap, \href
  {http://adsabs.harvard.edu/abs/1999A%26A...344..991P} {344, 991}

\bibitem[\protect\citeauthoryear{{Prikryl}}{{Prikryl}}{1979}]{prikryl1979}
{Prikryl} P.,  1979, Bulletin of the Astronomical Institutes of Czechoslovakia,
  \href {http://adsabs.harvard.edu/abs/1979BAICz..30..321P} {30, 321}

\bibitem[\protect\citeauthoryear{{Prikryl}}{{Prikryl}}{1983}]{prikryl1983}
{Prikryl} P.,  1983, Bulletin of the Astronomical Institutes of Czechoslovakia,
  \href {http://adsabs.harvard.edu/abs/1983BAICz..34...44P} {34, 44}

\bibitem[\protect\citeauthoryear{{Stober}, {Matthias}, {Brown}  \&
  {Chau}}{{Stober} et~al.}{2014}]{stober2014}
{Stober} G.,  {Matthias} V.,  {Brown} P.,   {Chau} J.~L.,  2014, \mn@doi [\grl]
  {10.1002/2014GL061273}, \href
  {http://adsabs.harvard.edu/abs/2014GeoRL..41.6919S} {41, 6919}

\bibitem[\protect\citeauthoryear{{{\v S}imek} \& {Pecina}}{{{\v S}imek} \&
  {Pecina}}{1999}]{simek1999}
{{\v S}imek} M.,  {Pecina} P.,  1999, in {Baggaley} W.~J.,  {Porubcan} V.,
  eds, Meteoroids 1998. p.~87

\bibitem[\protect\citeauthoryear{{{\v S}imek} \& {Pecina}}{{{\v S}imek} \&
  {Pecina}}{2002}]{simek2002}
{{\v S}imek} M.,  {Pecina} P.,  2002, \mn@doi [Earth Moon and Planets]
  {10.1023/A:1015816116844}, \href
  {http://adsabs.harvard.edu/abs/2002EM%26P...88..115S} {88, 115}

\makeatother
\end{thebibliography}








\bsp	
\label{lastpage}
\end{document}